\documentclass[pra,twocolumn]{revtex4}
\usepackage{graphicx}

\newcommand {\cD} {{\cal D}}
\newcommand {\cQ} {{\cal Q}}
\newcommand {\cP} {{\cal P}}
\newcommand {\cF} {{\cal F}}

\newcommand {\cH} {{\cal H}}

\newcommand{\con}{\setlength{\itemsep}{0cm}\setlength{\parskip}{0cm}}

\begin{document}
\title{ Nobel Lecture: Multiple equilibria}
\author{Giorgio Parisi}\email{giorgio.parisi@gmail.com}
\affiliation{Dipartimento di Fisica,
Universit\`a di Roma {\it La Sapienza},
\\
INFN, Sezione di Roma I, CNR-NANOTEC UOS Roma \\
Piazzale A. Moro 2, I-00185 Roma, Italy}

\begin{abstract}
This is an extended version of my Nobel Lecture, delivered on December $8$ 2021. I will recall the genesis of the concept of multiple equilibria in natural sciences. I will then describe my contribution to the development of this concept in the framework of statistical mechanics. Finally, I will briefly mention the cornucopia of applications of these ideas both in physics and in other disciplines.
\end{abstract}

\maketitle

\section{The interplay of disorder and fluctuations in physical
systems from atomic to planetary scales}
The Nobel Prize in Physics 2021 was awarded {\sl for groundbreaking contributions to our understanding of complex physical systems} with one half jointly to Syukuro Manabe and Klaus Hasselmann {\sl for the physical modeling of Earth's climate, quantifying variability and reliably predicting global warming} and the other half to Giorgio Parisi {\sl for the discovery of the interplay of disorder and fluctuations in physical systems from atomic to planetary scales}.

When I gave my 25 minutes Nobel lecture "{\sl Mulitple Equilibria}" I was unable to cover all the work that I have done on systems from atomic to planetary scale. This impossibility persists in this written version, also because I would like to avoid adding too much material to the oral version.
For this reason, I will not mention many very interesting subjects:
\begin{itemize} 
\item Random field Ising model and dimensional reduction.
\item Intermittency in turbulence and multifractals.
\item Stochastic interfacial motion (i.e. the Kardar-Parisi-Zhang equation).
\item Stochastic quantization.
\item The flight of starlings.
\end{itemize}

Other very important related topics I will briefly mention are:
\begin{itemize}
\item Stochastic resonance.
\item Non-equilibrium fluctuations.
\item Granular matter (hard spheres).
\item Random Laser.
\item Theoretical aspects of finite-dimensional spin glasses.
\item Large scale simulations of spin glasses.
\item Optimization theory, constraint satisfaction problems: 3-SAT, coloring, etc.
\item Neural networks.
\end{itemize}

A review of various developments of these ideas is contained in the one hundred authors' book {\sl Spin Glass Theory and Far Beyond – Replica Symmetry Breaking after 40 Years} \cite{THEBOOK}.

\section{The antefact}
In 1972 Niles Eldredge and Stephen Jay Gould proposed the evolutionary theory of punctuated equilibria \cite{ELDGOU1972}. In a nutshell the evolution of the species is not a continuous gradual process: there are long periods of stasis with practically no changes in the morphology and these periods are punctuated by rare bursts of evolutionary change. The stability of a system composed of many random components started to be an interesting mathematical issue \cite{MAY} in the same years.

Many other different systems have long periods of equilibrium separated by fast transitions to a new equilibrium point. This may happen for example in ecosystems, in climate (glaciations), geological eras, and so on. In the original stochastic resonance model for glaciations, the climate has two equilibrium states.

Generally speaking, we could say that a complex system can stay in many different equilibrium states, while a simple system may stay only in one or a few equilibrium states. For example, an animal like a dog can do many different actions (e.g. play, sleep, eat, hunt ...); an animal can switch from one state to another state in a very short time as the effect of a small perturbation, e.g. suddenly waking up after hearing a suspicious noise. 

Similar considerations can be done in Hebb's theory \cite{HEBB} of memory (1949) which was modelized with great success first by Little \cite{LITTLE} (1974) and finally by Hopfield \cite{HOPFIELD} (1982) with his very successful theory of associative neural networks. The part of the brain that is responsible for memory may stay in an extremely large number of different equilibrium states (attractors \cite{AMIT}), each corresponding to the recalling of a different item: it may remain in that situation for a long period, switching from one memory to the other as an effect of an external perturbation. The extremely large number of items that we can memorize and recall is related to the extremely large number of possible equilibrium states.

Similar ideas have been put forward by Goldstein \cite{GOLDSTEIN} in 1969 to understand the physics of standard structural glasses (like window glass): as explained in \cite{CavagnaPedestrian} {\sl Goldstein's idea is that at low enough temperatures, a supercooled liquid explores the phase space mainly
through activated jumps between different amorphous
minima, separated by potential energy barriers.}

I notice en passant that "{\sl multiple equilibria}" is a well-known concept to economists: for example in 1970 the Nobel Laureate Gerard Debreu wrote the paper Economies with a Finite Set of Equilibria \cite{DEBREU}. The title of the Nobel lecture in the economics of 2022 given by Philip Dybvig was "{\sl Multiple equilibria}": the same title in two consecutive years but in two different disciplines.

\section{Spin glasses: the starting point}%%%%%%%%%%%%%%%%%%%%%%%

The concept of systems with many equilibrium points was floating in different parts of science in the seventies, including physics. However, it was not clear how to attach the problem using the standard tools of physics. The systems were rather complex and so different from the others; people working in glasses did not know the {\sl punctuated equilibra} theory; moreover, such a piece of knowledge would be useless: it would have been too difficult to construct an appropriate model.

To make progress physicists need to find the simplest problem, model it, and understand the properties of the model in mathematical terms: only at a later stage, they can use the acquired knowledge as a trampoline for other problems.

\begin{figure}
\includegraphics[width=1\columnwidth]{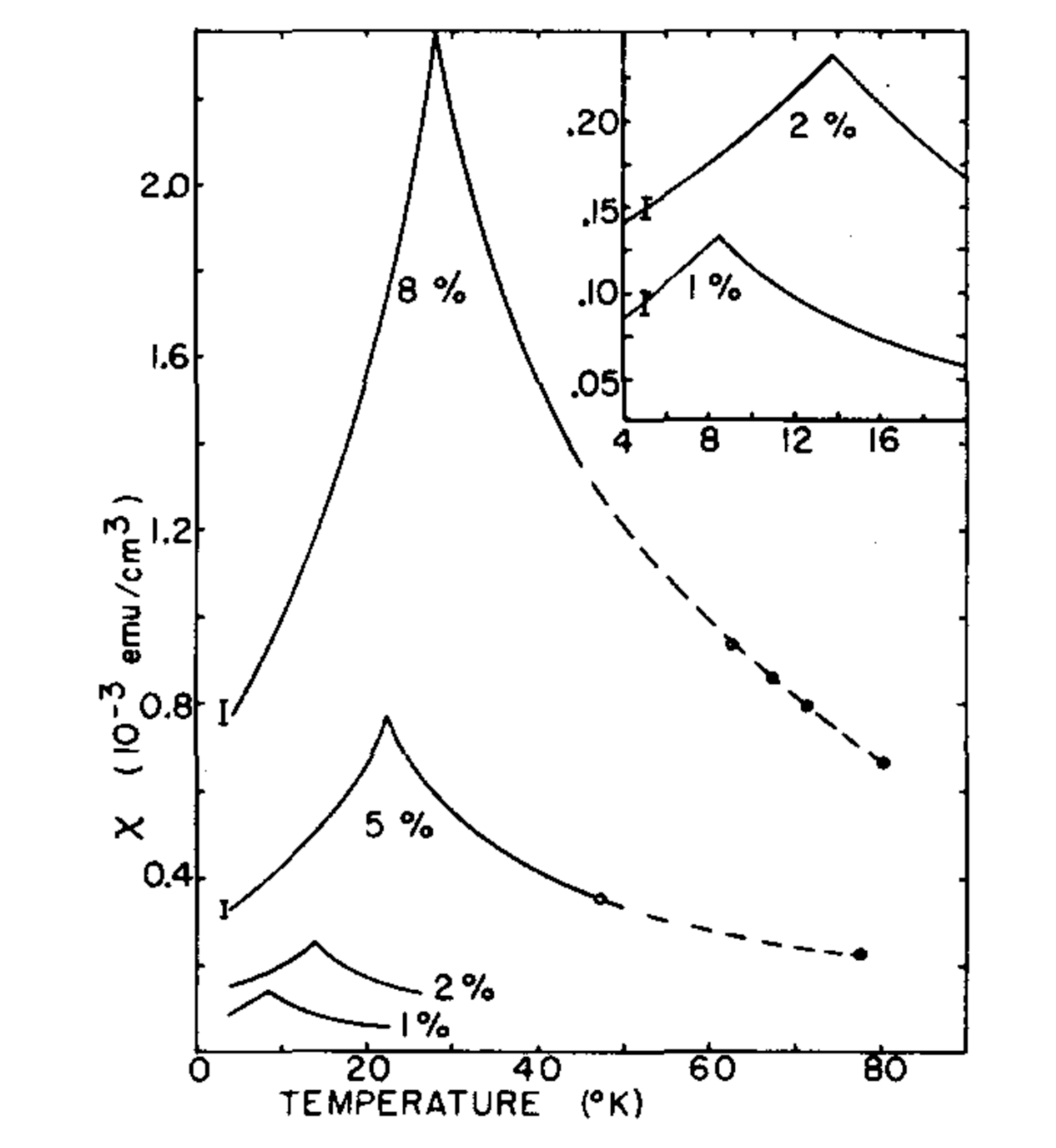}
\caption{ The magnetic susceptibility of spin glasses: the results for Au-Fe alloys in the region of low Fe concentration,  from \cite{Cannella}.}\label{CANNELLA}
\end{figure}

Things started to change unexpectedly in 1971. After a long series of experiments, it was proved \cite{Cannella} that an alloy of gold with a few percent of iron undergoes a phase transition: at a very small magnetic field, the magnetic susceptibility had a sharp pick, indicating a transition to some kind of antiferromagnetic order (see fig. (\ref{CANNELLA})). 

How do these materials behave? In standard ferromagnetic systems, a pair of spins decrease its energy when the two spins point in the same direction: this is what happens when two iron atoms are in contact.

In the spin glass case, the iron atoms are in random positions inside the gold matrix: at a low percentage of iron, they are not in contact and the interaction is mediated by the presence of the gold atoms. However, the sign of the interactions depends on the distance of the atoms of iron: sometimes the interaction wants to put a pair of spins in the same direction, and sometimes in an opposite direction. 

A triplet of spins is said to be frustrated if we cannot satisfy all the requests: for example, if spin $A$ want to be in the same direction as spin $B$, spin $B$ want to be in the same direction as spin $C$, and spin $C$ want to be in the opposite direction of spin $A$, we cannot satisfy all the requirements simultaneously because they are not compatible. 

In the same way, if John and Fred want to sit at the same table, Fred and Bob want to sit at the same table, but John and Bob hate to sit at the same table \cite{MPV,PaBook2} we have frustration as stressed by Toulouse \cite{Tou77} (and in real life a lot of discussions).
Of course, frustration, i.e. non-compatible requirements, is ubiquitous in the real world, from Shakespeare's tragedies to the problem of putting pieces of baggage in the trunk. Disorder and frustration are essential features of spin glasses.

Theorists started to work and models were constructed. Edwards and Anderson (1975) \cite {EA} arrived at the first simple model: the spins ($\sigma_i$) are located on a $D$-dimensional lattice (the index $i$ denotes the lattice point): their number $N$ is equal to $L^D$, $L$ being the side of a $D$-dimensional cube. In the simplest versions, the spins $\sigma$'s are Ising variables (they can take the values $\pm 1$). 

The Hamiltonian of the system is
\begin{equation}
H_J[\{\sigma\}]= -\frac12 \sum_{i,k=1,N} J_{i,k} \sigma_i \sigma_k +h\sum_i \sigma_i\,,
\end{equation}
where the couplings $J_{i,k}$ are different from zero only if the point $i$ and $k$ are nearest neighbors: they are independent random variables [Gaussian distributed or binary ($\pm 1$)] and $h$ is a constant magnetic field.

The paper is remarkable for very important progress:
\begin{itemize}
\item We expect that at low-temperature spin glasses develop a spontaneous magnetization: $m_i\equiv \langle \sigma_i\rangle$, where $\langle \cdot \rangle$ denotes the thermal expectation value. However, each sample is characterized by different $J$'s and it will have different values of the $m_i$. We face the problem of studying the statistical behavior of an {\sl ensemble} of different systems and new ideas were needed to reach this goal. 

Their great idea was to consider as an order parameter the quantity $q_{EA}\equiv \mbox{Av}(m_i^2)$, where $\mbox{Av}$ denotes the average over the system [i.e. $\mbox{Av}(m_i^2)\equiv N^{-1}\sum_{i=1,N}m_i^2]$.

 In principle, $q_{EA}$ could depend on $J$ but this possibility was correctly disregarded.
At zero magnetic field ($h=0$), the order parameter $q_{EA}$ is zero at temperatures above the critical temperature and it becomes positive at temperatures below the critical temperature. Naively the magnetic susceptibility $\chi$ is given by $\chi=\beta(1-q_{EA})$, so that the change of behavior in the magnetic susceptibility at the critical point can be interpreted as the appearance of a non-zero $q_{EA}$.

\item We face the problem of computing the average free energy $F$. Let me write down a few definitions:

\begin{eqnarray}
Z_J(\beta,N)=2^{-N}\sum_{\{\sigma\}} \exp \left(-H_J[\{\sigma\}]\right)\,,\nonumber\\
F(\beta)=-\lim_{N\to\infty}\frac{\overline{\log(Z_J(\beta,N))}}{\beta N} \, , \label{REP}
\end{eqnarray}
where the overline denotes the average over the $J$. To lighten the notation in the text we will not indicate the obvious dependence on $N$ and $\beta$ of these two quantities.

The average over the $J$ is rather complex: however, it can be done in a simple way using an outrageous trick.

We first notice that $Z_J^n$ is the product of $n$ factors $Z_J$. In this way $Z_J^n$ is the partition function of a system with $nN$ spins: each system of $N$ spins is replicated $n$ times. Using this representation we can do the average over $J$ and we get formulae that involve the integral over $n(n-1)/2$ variables.

The second step is to use the formula
\begin{equation}
\overline{\log(Z_J(\beta,N))}=\lim_{n\to 0}\overline{\frac{Z_J(\beta,N)^n-1}{n}}\, .
\end{equation}

\end {itemize}

It is clear that we are cheating: at the beginning, the number of factors ($n$) must be an integer (half a factor or $\pi$ factors do not make sense) and in the second step $n$ is a real number. However, physicists often do illegal manipulations, hoping that in the end, the results remain correct. 

This approach is called the replica method (or the replica trick) because we consider the partition functions of $n$ identical replicas of the same systems.

Equivalently we can also define
\begin{equation}
F_n(\beta)=-\lim_{N\to\infty}\frac{\overline{Z_J(\beta,N)^n}}{\beta n N}\,,\quad F(\beta)=\lim_{n\to 0} F_n(\beta)\,.
\end{equation}
Once the free energy $F_n$ is computed for integer $n$ (in the infinite $N$ limit), we analytically continue $F_n$ to $n=0$ in order to get the average free energy $F$. 

The function $F_n$ is well defined for non-integer $n$, however, this naive approach fails if the function $F_n$ is not analytic in $n$ and it has a singularity (i.e. a point of non-analyticity) for non-integer $n$ (e.g. for $0<n<1$). Such a singularity can be seen only in the infinite $N$ limit.

A subsequent simplification was taken by Sherrington and Kirkpatrick \cite{SK}: the Hamiltonian has the same form as before with the difference that all pairs of the spins have a direct interaction; the average of the couplings is given by $\overline{J^2}=1/N$ where $N$ is the total number of spins. 

According to the folklore in statistical mechanics, when all the components have a direct infinitesimal interaction in the limit $N\to\infty$, each component feels only the collective behavior of the system: physical space is no more present and the correlations have an infinite range. Such a model should be soluble, in the sense that one can arrive at a closed set of equations in the infinite volume limit. However, the solution is correct only if we have correctly identified the collective behavior of the system. 

The solution of the infinite range model should provide the useful mean field theory of the model: the large dimension limit of the lattice-defined theory should be described by this mean-field theory, as has been proved in many cases.

The paper \cite{SK} was a turning point: the goal became to obtain the solution of the SK model.
The title {\sl Solvable model of a spin-glass} was correct, the model is actually solvable, but the paper did not contain the correct solution because, as stressed by the authors, the reported solution of the model produced a negative value of the entropy at low temperature: it was an internal contradiction because the entropy by definition is non-negative. Some subtle mistake was made.

The derivation of SK was simple. After some computations based on Gaussian integrations, it was possible to write exactly for a system of $N$ spins:
\begin{equation}
\overline{Z_J(\beta,N)^n}=\int dQ \exp (-N\beta F_n(Q)) \approx \exp \left(-N\beta F_n(Q_n^*)\right), \label{INT}
\end{equation}
where $Q$ denotes a symmetric $n\times n$ matrix that has diagonal elements equal to zero; the integral runs over all these matrices. The function $F_n(Q)$ has an explicit form that it is not worthwhile to write here and $Q_n^*$ is the value of $Q $ that {\sl minimizes} $F_n(Q)$. The simplest option for finding candidates for a minimum is to look for the solutions of the stationary point equation $\partial F_n(Q)/\partial Q=0$;

Now Sherrington and Kirkpatrick worked directly at $n=0$; they made also an extra innocent-looking assumption
\begin{equation}
Q_{a,b}^*= q, \ \mbox{for} \ a\ne b \,. \label{SYM}
\end{equation}
This assumption was quite natural: both the integrand and the integral in eq. (\ref{INT}) are invariant under the reshuffling of the $n$ indices (i.e. the action of the permutation group of $n$ elements, $S_n$, called also the replica group).

In a nutshell this symmetry stems from the commutativity of the multiplication: when we write $Z_J^n$ as the product of $n$ factor $Z_J$, we attach a label $a$ ($a=1,n$) to each of these factors, but the order in which we perform the multiplication is irrelevant. The matrix $Q^*
$ in eq. (\ref{SYM}) is the only one that is symmetric under the action of the replica group, so it is the obvious choice.

After two years De Almeida and Thouless proved \cite{DEATHO} that this innocent-looking assumption was wrong: the value of $Q^*$ chosen by SK was not a {\sl minimum}; more precisely the replica symmetric matrix of eq. (\ref{SYM}) is a minimum for $n>n_c$, but it no longer a minimum for $n<n_c$, with $0<n_c<1$. So the worst-case scenario of the non-analyticity of the function $F_n$ was realized. 

The reader should notice that the precise definition of {\sl minimum} is unclear in the context. However, everybody agreed that a stationary point of the function $F$ (i.e. a point such that $\partial F /\partial Q_{a,b}=0$) is a candidate to be a minimum only if the Hessian matrix of second derivatives 
\begin{equation}
\cH_{a,b;c,d}=\frac{\partial^2 F}{\partial Q_{a,b}\partial Q_{c,d}}
\end{equation}
is non-negative, (i.e. its eigenvalues are non-negative). This statement generalizes the well-known one-dimensional result that if $f'(x^*)=0$ and $f''(x^*)<0$ the function $f(x)$ has a maximum at $x=x^*$, certainly not a minimum.

The subtle mistake was found by de Almeida and Thouless, but, unfortunately, these authors were not able to suggest which form could have the {\sl minimum}; it was clear the saddle point matrix $Q^*$ could not be left invariant by the action of the replica group: the replica symmetry should be spontaneously broken, but the space of non-symmetric matrices is very very large. 

A natural possibility was that the correct saddle-point matrix $Q^*$ is symmetric under a subgroup of the replica group $S_n$, but also the number of different subgroups is very large, especially in the slippery limit $n\to0$. Another possibility was to abandon the replica method but it was unclear what method to use in place of the replica method.

A few attempts to find the correct minimum were done, but they were unsuccessfully \cite{13,14}. Important progress on the physics of the problem was made  in the TAP paper (Touless, Anderson, and Palmer \cite{TAP}), where it was shown that the low-temperature behavior of the model was very different from that of SK. More results in this direction were contained in the very influential Les Houches lecture notes by P.W Anderson \cite{ILL}.

\section{The exact solution of the model.}

The history of the solution of the model and the consequent discovery of replica symmetry breaking is quite complex. The interested reader can find a detailed exposition in the monumental and choral work \cite{DZ,HIST}. Here I will expose mainly the history from my personal viewpoint, which is exposed at length in my interview in \cite{DZ}.

My interest in the problem started at the end of 1978. I was working with Sourlas and Drouffe on lattice QCD in high dimensions \cite{DPS}. In our formalism, the corrections to mean field theory for lattice gauge theories were related to the properties of interacting branched polymers.

Looking at the literature in the library, I encountered the replica method in the study of branched polymers \cite{LUB} and I learned that the replica method was giving incorrect results in the case of the SK model. I immediately wanted to understand why. I started to study, to look for a different form of the matrix $Q^*$, but none were satisfactory.

After many trials I had an intuition: in other papers, \cite{13,14}, the $n$ indices were divided into $n/m$ groups of $m$ elements each and the value of the matrix elements of $Q$ depends on the group. Everybody was assuming that $m$ was an integer: Blandin and collaborators \cite{13} were using $m=2$; De Dominicis and Garel \cite{DG} generalized this approach to a generic integer $m$ and considered also the $m\to\infty$ limit: in this limit, the entropy problem was solved, but other low-temperature properties were not in agreement with the simulations either with the theoretical analysis of TAP.

I made the bold assumption that $m$ could be a non-integer number, more precisely a number in the interval [0-1]. For example, I was dividing the $n$ replicas into $2n$ groups of $1/2$ replicas each.
Of course, that is crazy, but my viewpoint was I should first check if this crazy idea was leading to correct results and postpone other questions to a later stage. It was clear to me that also non-integer angular momentum is a crazy idea, but this idea led to Regge poles.

This approximation is what is now called one-step replica symmetry breaking. Everything was nearly perfect. The theory predicted values of the specific heat and the energy in agreement with the numerical simulations. The entropy remained a problem, but instead of being $-0.16$ at zero temperature, it was $-0.01$: still negative but quite near zero. I wrote the results in a short note \cite{GP79} that I submitted to Physics Letters A.

At the end of the paper, I added the observation that one could improve the theory by dividing the $n/m$ groups into $(n/m)/m_1$ groups of replicas, where $m_1$ was a new variational parameter. I was also conjecturing the correct solution was obtained when the procedure was repeated an infinite number of times. 

Some fancy group theory arguments also were added, arguing the permutation group of zero objects is an infinite group because it contains itself as a proper subgroup. Indeed the subgroup of $S_n$ that does not change the matrix $Q^*$ is the semidirect product of the permutation group of $n/m$ elements $S_{n/m}$) with the direct product of $n/m$ copies of the permutation group of $m$ elements ($S_m$). For $n=0$ also $n/m=0$, so that $S_0$ contains as a subgroup the semi-direct product of $S_0$ with the direct product of zero factors $S_m$. This sentence looks like a nonsensical text written by modern AI.

The response of the referee was remarkable. In a nutshell: {\sl The approach does not make sense, but the numbers coming from the formulae are reasonable, so it can be published. The last observation is not worth the paper on which it is written and it should be removed}. 

I laughed because in the meanwhile I extended the computation to an infinite number of subdivisions: I found that I could associate a continuous function $q(x)$ (where $x$ belongs to the interval $[0-1]$ to the matrix $Q$ when one repeats the process of symmetry breaking an infinite number of times. 
In other words, I arrived at the formula
\begin{equation} 
F=\max_{q(x)} F[q(x)] \,,
\end{equation}
where $F[q(x)]$ is a functional of the function $q(x)$. I first computed it near the critical temperature \cite{GP79b} and finally at all temperature \cite{GP79c}.

At the end of 1979, I got the expression for the free energy of the model. At zero magnetic field, we have

\begin{equation}
F[q(x)]=-\frac{\beta}{4}\left[ 1+ \int_0^1 \, dx\, q^2(x) -2q(1)\right] - \frac{1}{\beta}f(0,0) \,, \label{freeRSB}
\end{equation}
where $f(x,h)$ is an auxiliary function defined in the strip $0\le x\le1$ and it is the solution to the non-linear antiparabolic equation 
\begin{equation}
\frac{\partial f(x,h)}{\partial x}=-\frac{1}{2} \frac{{\rm d} q}{{\rm d} x} \left[ \frac{\partial^2 f}{\partial h^2} + x \left( \frac{\partial f}{\partial h} \right)^2\right]\,,
\end{equation}
with the initial condition 
$f(1,h)=\ln (2 \cosh (\beta h))$.
The zero temperature entropy was zero, as it should be.

It is remarkable that in the final formula, the craziness disappeared: we come back to respectable mathematics with one big surprise: in all known mean field theory the order parameter was a number, here it is a function \cite{MPV,parisibook2}.

But what is the physical meaning of the function $q(x)$? How could we derive the relatively simple final formulas without using the nonsensical mathematics described before? Most importantly are the final formula for the free energy correct? 

At that time there was no suggestion of how this result could be rigorously proved: the very existence of the free energy density in the large $N$ limit was lacking rigorous proof. The difficulty was to prove that all the random systems have the same free energy density when $N$ goes to infinity: we needed a clever generalization of the law of large numbers.

There were some hints on the meaning of having a non-constant $q(x)$. If $q(x)$ were constant, i.e. $q(x)=q$, we would recover the original SK solution.
By an ingenious trick (adding an infinitesimal term that breaks the replica symmetry) in \cite{13} it was argued that when replica symmetry was broken, $q_{EA}$ is given by the largest matrix element of the matrix $Q$. In this context, this result leads to:
\begin{equation}
q_{EA}=\max_x q(x)\,, \quad \chi=\beta\int_0^1 dx (1-q(x))\,.
\end{equation}
Unfortunately, a back-of-the-envelope calculation gave $\chi=\beta(1-q_{EA})$ which was not the right result. These formulas and this discrepancy were suggesting something, but what?

Despite this hint, I had no sound idea of the meaning of $q(x)$ and I was unable to make further advances. At the end of 1979, I started to study other problems, waiting for inspiration for further progress.

The belief in the correctness of the approach was strongly reinforced by two papers. In the first paper \cite{TAK} it was shown that near the critical temperature, the values of negative eigenvalues of the Hessian in the one-step formulation were reduced by a factor of 9 indicating that one was moving in the correct direction. Finally in a later and conclusive paper \cite{DK0} it was shown that in the formulation with an infinite number of replica symmetry breaking, the negative eigenvalues of the Hessian are no more present: one remains with a rather intricate structure of zero modes and nearly zero modes.

\section{The discovery of complexity, with the help of my friends}
In the summer of 1982, I lectured at the Les Houches summer school on spin glasses \cite{LH} and in the fall I went to Paris, at the IHES, for a two-month visit. I had many discussions about spin glasses which was a hot subject: and a lot of progress was made. People had already realized that spin glasses could be in many stationary states \cite{Bray,DGGO}, most of them having higher free energy (i.e. metastable equilibrium states). However, the connection of this fact with the replica symmetry-breaking solution described above was not clear \cite{DY}. 

In the end, I found that the replica symmetry-breaking solution implies that a spin glass may have many equilibrium states \cite{GP83}, labeled by an index $\alpha$ ($\alpha=0,1\cdots \infty$): each of these states appears at equilibrium with a probability $w_\alpha$. 

Each state is characterized by its own magnetizations $m_i^\alpha$. One can define the matrix of overlaps 
\begin{equation}
\cQ_{\alpha,\gamma}=\frac{ \sum_i m_i^\alpha m_i^\gamma}{N}\equiv \mbox{Av}(m_i^\alpha m_i^\gamma)\,.
\end{equation}
The Edwards Anderson order parameter ($q_{EA}$) is given by $\cQ_{\alpha,\alpha}$.

Each sample has its own matrix $\cQ$ and its own weights $w$. In other words for each sample, we could define a descriptor $\cD_J$ that is formed by the matrix $\cQ$ and the weighs $w$'s.

To describe the connection with the function $q(x)$, it is convenient to consider the probability of finding two states with overlap $q$ in a given sample: for very large, but finite, $N$, it is given by
\begin{equation}
P_J(q)=\sum_{\alpha,\gamma} w_\alpha w_\gamma \delta\left( \cQ_{\alpha,\gamma} -q\right)\,.
\end{equation}
The function $P_J(q)$ depends on the system. The simplest quantity to consider is its average over the samples:
\begin{equation}
P(q)=\overline{P_J(q)}.
\end{equation}

The connection with my theory was surprising: if we define 
\begin{equation}
x(q)= \int_0^q dq' P(q') \,,
\end{equation}
one could prove that $x(q)$ is the inverse of the function $q(x)$ that I introduced in the solution using replicas.

The interpretation of $q(x)$ was now clear. The quantity $x$ is a probability: indeed it is the probability that two replicas of the system have an overlap less than or equal to $q$. This probabilistic interpretation of $x$ explains why $x$ is in the range from zero to one.

The study of the statistical properties of the function $P_J(q)$ and of the descriptor $\cD_J$ were done together with very good friends of mine \cite{5a,5b}. The results were so unexpected that when I saw the first results of our computations, my first thought was {\sl my theory must be wrong if it produces such nonsense}. 

\begin{figure}
\includegraphics[width=0.99\columnwidth]{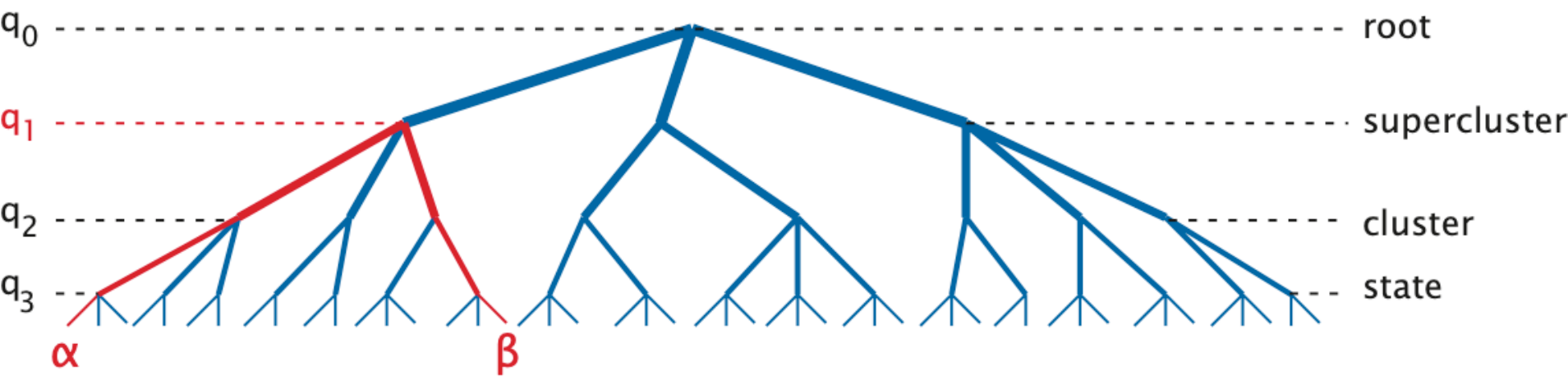}
\caption{An example of the tree of states in the case where the overlap takes only three possible values: taken from \cite{PRY}. The reader should notice that the tree is upside down: the root is at the top and the leaves are at the bottom. }
\label{Tree}
\end{figure}

After a lot of work, we realized that the predictions made a lot of sense:
\begin{itemize}
\item The probability distribution of the overlap changes from system to system also in the infinite volume limit. This is quite amazing because usually intensive quantities do not fluctuate. However, the self overlaps $\cQ_{\alpha,\alpha}$ do not depend on $\alpha$ and on the system. It is possible to show that $\cQ_{\alpha,\alpha}=q_{EA}$.
\item The probability distribution of the $w$'s follows {\sl simple} mathematical rules that I will not describe here.
\item The number of equilibrium states is infinite. This was the first case of an infinitely numerable number of different equilibrium states. This is a quite strange situation because it violates the Gibbs rule.
\item The most surprising result was that in presence of an arbitrarily small non-zero magnetic field, the states satisfied the ultrametricity property
\begin{equation}
d_{\alpha,\gamma}\le \min(d_{\alpha,\beta},d_{\beta,\gamma})\quad \forall \beta\,, 
\end{equation}
where the distance between two states is defined in a natural way:
\begin{equation}
d_{\alpha,\gamma}^2=\cQ_{\alpha,\alpha} +\cQ_{\gamma,\gamma}- 2 \cQ_{\alpha,\gamma}= \mbox{Av}
\left(m_i^\alpha -m_i^\gamma\right)^2\,. 
\end{equation}
Without entering into the mathematical details, ultrametricity implies that the states can be assigned to the leaves of a tree and that the distance between states is proportional to the height you have to climb the tree for going from one leaf to another leaf. 

An explicit probabilistic construction of this tree is given in \cite{PRY}, while a rigorous construction of the tree was done in \cite{Ruelle}. In fig. (\ref{Tree}) I show an example, where only a finite number of the branches are depicted, i.e. those with higher probability.

A model that by inspection has the same distribution of weights of the one-step replica broken case is the Random Energy Model of Derrida \cite{REM} that generalizes to the GREM model for more than one-step replica symmetry breaking \cite{GREM}.
\end{itemize}

Ultrametricity and taxonomy are essentially related: a standard taxonomy, i.e. a hierarchical classification, is possible only if the relevant properties have an ultrametric structure. In the standard taxonomic classification of living beings, the distance is related to the history of evolution. In spin glasses, the taxonomy is intrinsic to the static equilibrium properties of the system and it is not related to evolution in time.

In spin glasses and other systems, where the function $q(x)$ contains a continuous part with a non-zero derivative, the function $P(q)$ is non-zero in an interval. This situation is called continuous replica symmetry breaking. Here the tree of the states is branching at any level: in any height interval, there is an infinite number of branches. It is an extremely complex construction that appeared unexpectedly. The presence of such construction for the probability of the descriptors $\cP(\cD)$ reflects the extreme complexity of the rugged landscape of spin glasses.

In other cases the function $P(q)$ is the sum of two or more delta functions: for example, if the function $P(q)$ contains only two delta functions, we have the so-called one-step replica symmetry breaking, which was the first case we  considered. When there are three deltas, we have two steps replica symmetry breaking (a very rare case). Finally, when only one delta function is present we are in the usual case where there is no replica symmetry breaking.

The crazy form of the $Q$ matrix is equivalent to this extremely complex probability distribution of the descriptor. Indeed it was shown \cite{MPVART} that the formula for the free energy can be derived using probabilistic arguments starting from the assumption that the probability of the descriptors is just the one given by the replica formula.

\section{Mathematicians come to the rescue}

At this stage, at the beginning of the nineties, we faced an impasse. Numerical simulations for the SK model were in agreement with the predictions of the replica symmetry broken approach \cite{BY}. However it was possible that the correct form of the matrix $Q$ was different and more complex, and the correct form of the probability of the descriptors [$\cP(\cD)$] was much more intricate. It was difficult to conclude in a definite way.

A possible approach could be to consider a free energy that depends on $\cP(\cD)$, i.e. $\cF[\cP]$. We could conjecture that the correct free energy is given by
\begin{equation}
F=\max_\cP F[\cP] \label{MAX}\,.
\end{equation}
Unfortunately, there was no idea of how to prove the previous formula. Why "$\max$" and not the usual "$\min$"? How to justify this strange choice? 

Moreover, the space of the descriptors is infinite-dimensional and the space of possible functions on an infinite-dimensional space is extremely large. How can we explore such a big space? The very complicated ultrametric $\cP(\cD)$ of the replica approach is just the simplest example of a very large family of distributions.

I was very skeptical about the possibility of finding proof that all the assumptions done were correct. Fortunately, the situation changed rapidly. 

In 2002 Guerra \cite{GT} introduced a new mathematical technique (Guerra interpolation) that allowed him to prove that $F\ge F_R$, where $F_R$ is the free energy computed via the replica formula \cite{GuerraI}, vindicating the use of "$\max$". A few months later Talagrand \cite{Tala} beautifully twisted Guerra's argument and he proved that $F\le F_R$, arriving at the conjectured result $F=F_R$.

Unfortunately, this sophisticated proof did not have direct implications on the form of $\cP(\cD)$. The problem was solved differently.
It was shown that equation (\ref{MAX}) has a precise mathematical meaning \cite{ASS}. Moreover
Ghirlanda and Guerra \cite{GuerraSS} proved some identities (the GG identities) that restricted very strongly the form of $\cP(\cD)$ (see also \cite{ACON}).

In a longly-awaited paper after many partial results, Panchenko \cite{PANU} was able to show that the GG identities implied ultrametricity. There is only one possible form of $\cP(\cD)$ that satisfies the GG identities and this form is parametrized as a functional of the function $q(x)$. Using the Panchenko results one obtains directly to replica formula after some nowadays standard computations. This second proof \cite{PanBook} is more connected to physical intuitions.

Unfortunately at the present moment, nobody is able to transform the original approach with replicas into rigorous proof. There are some ideas on a possible approach \cite{CPV, PBOOK}, however, it is still not easy to transform these half-baked ideas into a mathematical proof. It seems to me \cite{PBOOK} that there must be some relevant deep mathematical theorems that have not yet been proven.

In the nutshell, the work of the mathematicians has been fundamental to clarifying the physical hypothesis at the basis of the replica approach and to a any possible doubts about the correctness of the results of this approach.

\section{Marginal Stability}

We have seen that the existence of multiple equilibrium states is the distinctive hallmark of glassiness \cite{MPV,parisibook2}. While at finite temperatures these equilibrium states may be metastable (or maybe not, depending on the system), the issue of metastability becomes irrelevant in the zero temperature limit. 

A crucial question is how these states differ one from the other and how they are distributed in the phase space (the phase space becomes an infinite dimensional space in the thermodynamic limit). Moreover, in this situation a small perturbation produces a large rearrangement of the relative free energy of the equilibrium states: after a perturbation, the lowest equilibrium state may be quite different from the previous one. The linear response theorem has to be modified to take care of the existence of many equilibrium states.

Two main possible scenarios have very different properties as discussed for example in \cite{REVISITED}. 

\begin{itemize} \con
\item Different equilibrium states are scattered in phase space in a nearly random fashion: they stay at a fixed non-zero distance from the others \cite{CGP1}. In this situation, if you live inside one equilibrium state, you do not feel the existence of other equilibrium states: the barriers are very high \cite{MPV}. We can call this scenario the {\sl stable glass}: it corresponds to one-step replica symmetry breaking. In mean field theory the barriers for going from one state to another are exponentially large in the system size \cite{CGP2,CGP3}. Beyond mean field theory they are expected to be exponentially large in the appropriate parameter.
\item The distribution of equilibrium states in phase space is very far from a random one. Each state is surrounded by a large number of other equilibrium states that are arbitrarily near to that given state: 
\begin{equation}
\min_{\gamma\ne \alpha}d_{\alpha,\gamma}=0\,.
\end{equation} 
One could say the equilibrium states form a kind of fractal set. The barriers for
 going from one state to another are much smaller than in the previous case. 

If you live in one equilibrium state, you feel that you can move in many directions (typically in the direction of nearby states) without increasing too much the free energy: in other words, there are nearly flat directions in the potential exactly like at a second order transition critical point \cite{MPV}. 

The spectrum of small oscillations within an equilibrium state has an excess of low-frequency modes that in some cases may dominate the one coming from other more conventional sources (e.g. phonons or magnons). The precise form of this anomalous low-frequency spectrum and the localization properties of the eigenvalues crucially depend on the details of the theory. 

These glassy systems are self-organized critical systems. We call this scenario the {\sl marginally stable glass} that corresponds to continuous replica symmetry breaking. (I notice en passant that expression {\sl marginal stability} was also used to characterize some off-equilibrium metastable states \cite{KWT} with a slightly different meaning). 

\end{itemize}

For reasons of space, I will consider here only the case of the marginally stable glass.

A great step forward in the study of marginal stable glasses was taken by De Dominicis and Kondor \cite{DK1}. They computed the matrices elements of the resolvent of the Hessian ($1 /(\cH +\lambda)$): in the framework of the mean field approximation, if we put $\lambda=p^2$, we obtain the correlation functions of the spins in momentum space. These correlations (in position space) are defined as 
\begin{equation}
\langle \sigma_a(x)\sigma_b(x)\sigma_c(0)\sigma_d(0)\rangle\equiv C_{a,b;c,d}(x)\,.
\end{equation}
Unfortunately, they depend on four indices and this makes the whole analysis quite complex.

In mean field theory, when replica symmetry is broken continuously these correlations are long-range with a power law decay that has a very complex structure as a function of the four replica indices. Depending on the replica indices, in Fourier space we have the following possible singular behavior at small momentum $p$:
\begin{equation}
A_2/p^2\,,\qquad A_3/p^3\,,\qquad A_4/p^4\,,
\end{equation}
the $A$'s being constants that have been evaluated.

The existence of long-range correlations (although we do not know the exponents outside mean field theory) is one of the fundamental predictions of continuous replica symmetry breaking and marginal stability. 

Fortunately, there are very clear experimental observations of a large correlation length in spin glasses \cite{ORBACH} that increases with the age of the system. These experiments have also been reproduced {\sl in silico} in great detail \cite{JANUSCORR,JANUSCORR1,JANUSCORR2}.

The De Dominicis-Kondor computation (a real tour de force) was done in the framework of mean field theory, but it underlines a general relation: continuous replica symmetry breaking implies marginal stability and power-decaying correlations at large distances. The detailed results should be valid in high dimensions. They clearly show the existence of long-range correlations that decay at a large distance with a dimensional-dependent exponent: the correlation length is infinite.

For not-too-high dimensional spin glasses, the situation is less clear, also because we have a limited command of the perturbation theory around the mean-field theory. The computations are technically very demanding: they have never been done without approximations, but I hope that they will be done in the future.

I believe that one should better understand Ward's identities related to replica symmetry breaking \cite{W1,W2}. Ward identities are the right tools to address cancellations due to symmetries.
The theoretical situation in a magnetic field is still worse because we do not know the upper critical dimension \cite{BrayR}, i.e. the dimension at which the exponents start to be non-trivial, although there is a recent suggestion that the upper critical dimension is 8 \cite{D8}.

The sad outcome is that we do not know at which dimensions the spin glass phase disappears. We have experimental and numerical evidence that there is a transition at zero magnetic field in three dimensions and that the transition is absent in two dimensions. It was suggested theoretically that the transition temperature goes to zero in dimension $D_c=2.5$ \cite{FPV}. Obviously, we cannot (for the moment) do numerical simulation in non-integer dimensions, however, we can resort to extrapolation or interpolation of critical quantities as a function of the dimensions of the space.

The value $D_c=2.5$ is in very good agreement with numerical {\sl interpolation} of the zero temperature exponents for the Edward Anderson model (the value $D_c=2.4986$ is estimated in \cite{Bo}) (a much less precise {\sl extrapolation} of the critical temperature \cite{PR3} gives $D_c\approx 2.65$). Numerical estimations for the free energy barriers \cite{MP3} in dimensions 3 are in good agreement with the approximation that predicts $D_c=2.5$.

When the magnetic field is zero, numerical experiments in dimension 3 or higher are in very good and detailed agreement with the qualitative predictions of mean-field theory. For example in finite dimensions marginal stability predicts the existence of long-range correlations \cite{DK1} that are clearly observed in many simulations (e.g. \cite{Janus}) together with many others predictions \cite{CINQUE}. One can also measure numerically the correlation functions that are related to the presence of ultrametricity in dimension 3 \cite{MPY}.

More than ten years ago it was suggested that also in some two and three-dimensional structural glasses, in particular, hard-sphere systems at infinite pressure (i.e., in the jamming limit), are marginally stable, in the restricted sense that they are unstable under an infinitesimal perturbation \cite{WYART0,WYART1,WYART2,LN}. 

Following an original idea of Kirkpatrick and Wolynes \cite{KW}, quite recently \cite{PZR,LARGED1,FINAL,BOOKPUZ}, as discussed in detail in the next section, a mean field model of hard spheres has been constructed and solved in the infinite-dimensional limit ($D\to\infty$, where $D$ is the dimension of the space where the spheres move). The model has many features similar to the SK model: all the stigmata of marginal stability are present here, suggesting that a similar situation also holds for some finite-dimensional glasses.

\section{Some more experimental and numerical confirmations}

Although experiments are the ultimate source of confirmation of a theory, numerical simulations have proven to be a remarkable tool. They are crucial for studying quantities that are not accessible experimentally and for quickly disproving wrong theories.

Indeed the core prediction of the replica theory is the existence of multiple equilibrium states that have the same macroscopic properties but they differ microscopically. Unfortunately at the present moment, it is impossible to measure simultaneously the value of a large number of spins in experiments, so this key property may be directly observed only in simulations. 

Another reason for the importance of simulations is the possibility to explore the behavior of the system in a short time window, that is not accessible to experiments: in this way, simulations are complementary to experiments.

\subsection {Spin glass susceptibilies}
The magnetic susceptibility measures how the magnetization changes by adding a magnetic field. However, in the low-temperature region, the magnetization depends on the protocol we use to thermalize the system and to add the magnetic field (a form of hysteresis): consequently, we can define protocol-dependent (or history-dependent) magnetization. 

A clear prediction of the theory is the existence of two susceptibilities in two extreme cases:
\begin{itemize}
\item When we add a small magnetic field and we force the system to remain in the same state, we measure the linear response susceptibility that is given by $\chi_{LR}=\beta(1-q_{EA})$.

\item When we add a magnetic field and we allow the system to jump to the thermodynamically favored state, we measure the thermodynamic susceptibility that is given by $\chi_{eq}=\beta \int dx (1-q(x))$.
\end{itemize}

\begin{figure}
\includegraphics[width=0.99\columnwidth]{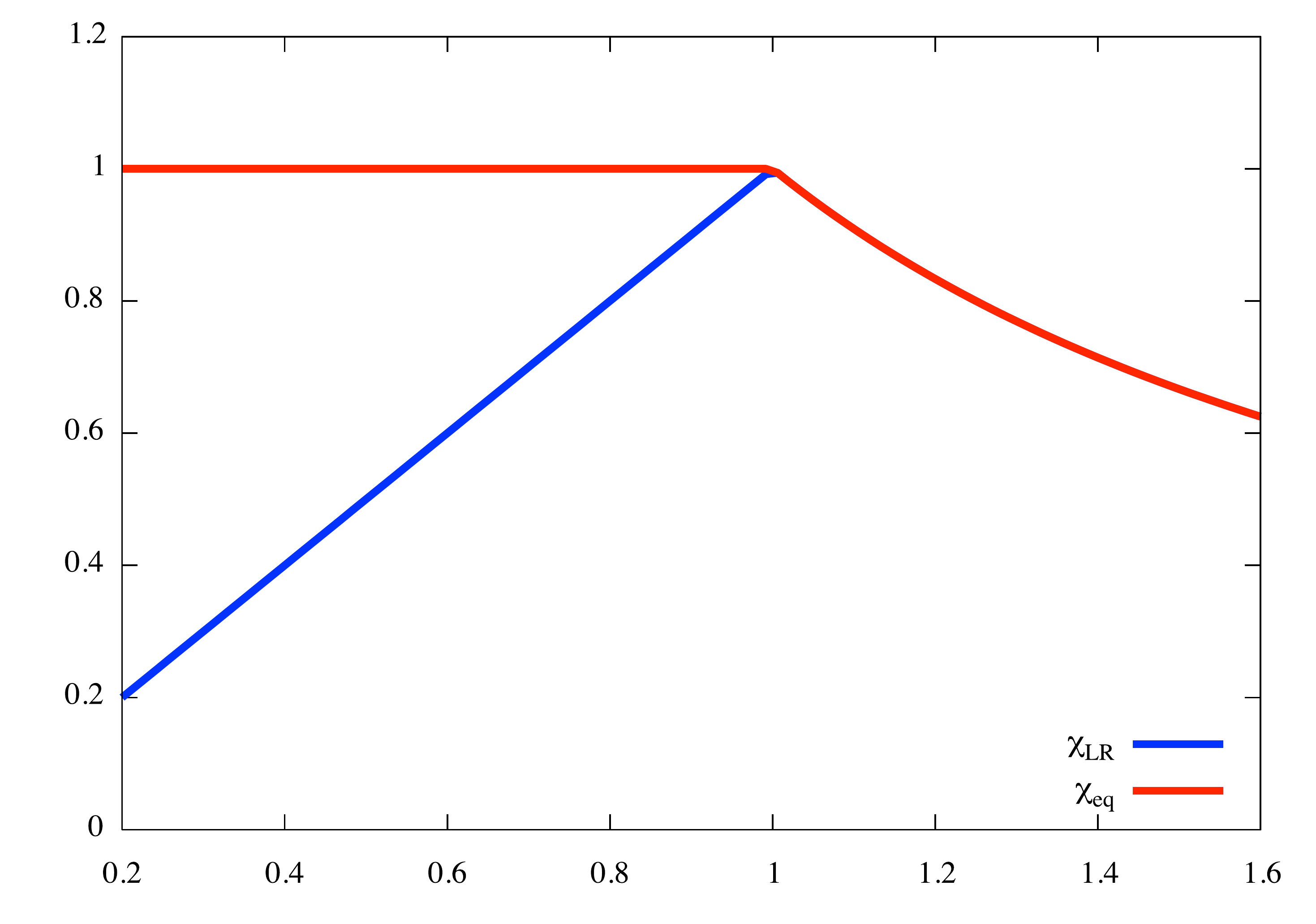}
\caption{The analytic results of the mean-field approximation for the linear response susceptibility ($\chi_{LR}$, lower curve) and the field cooled susceptibility ($\chi_{eq}$, upper curve). They coincide in the high-temperature region. \label{One} }
\end{figure}\begin{figure}
\includegraphics[width=0.99\columnwidth]{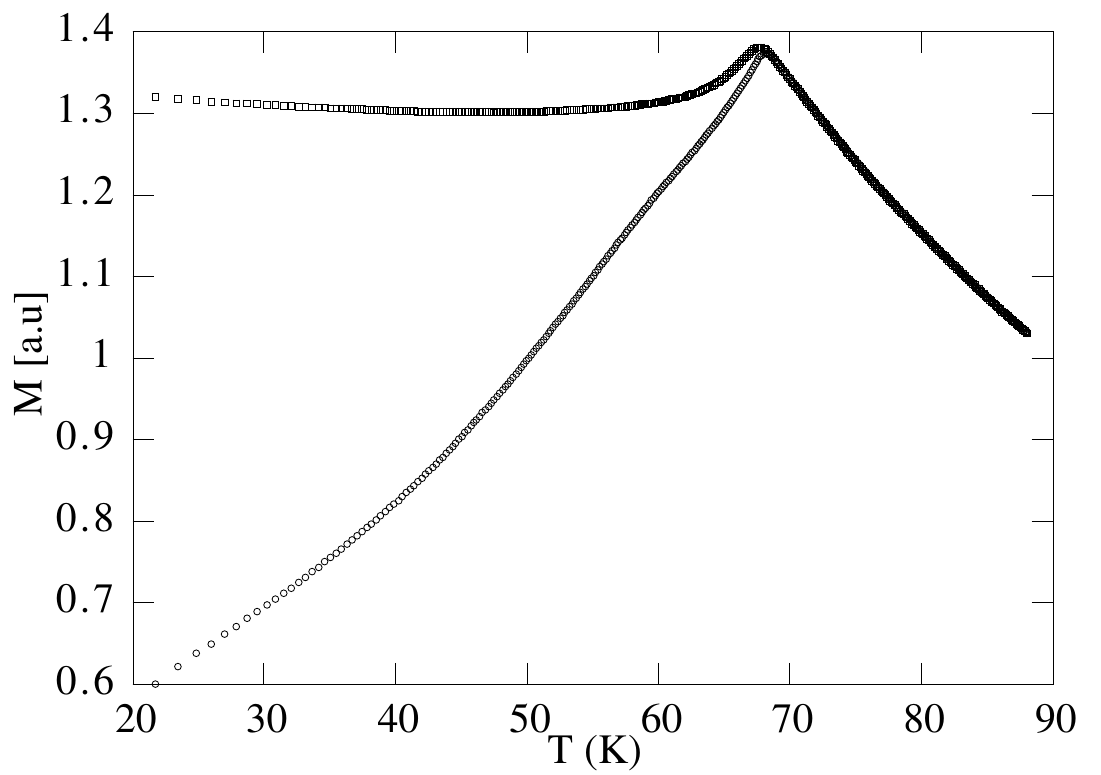}
\caption{The linear response susceptibility ($\chi_{LR}$, lower curve) and the field cooled susceptibility ($\chi_{eq}$, upper curve). The experimental results are taken from \cite{DJN}}.
\label{Two}
\end{figure}

The two susceptibilities have been measured experimentally in spin glasses and depicted in figures (\ref{One}) and (\ref{Two}) (mean-field theory and experiments respectively). 

The experimental protocols we consider are the following:
\begin{itemize}
\item The linear response susceptiibly ($\chi_{LR}$) is measured by adding a very small magnetic field when the system is already at the final low temperature. This extra field should be small enough to neglect non-linear effects. In this situation, when we change the magnetic field, the 
system remains inside a given state and it is not forced to jump from one state to another state: 
this is the ZFC (zero-field cooled) susceptibility, which corresponds to $\chi_{LR}$. 
\item
The second susceptibility ($\chi_{eq}$) can be approximately measured by cooling the system in presence of a small magnetic
field and comparing the observed magnetization with the one measured without this small magnetic field. In this case, the system has the ability to choose the state that is most appropriate in presence of the applied 
field. This susceptibility, the so-called FC (field-cooled) susceptibility, is experimentally nearly independent from the cooling rate. The quasi-independence of the field-cooled magnetization on the cooling rate confirms that the field-cooled magnetization is near to the equilibrium one: it is considered to be a good proxy of $\chi_{eq}$.
\end{itemize}

Summarizing, one can identify $\chi_{LR}$ and $\chi_{eq}$ with the ZFC susceptibility and with the FC susceptibility 
respectively. 

The theoretical plot of the two susceptibilities is shown in fig. (\ref{One}). As we discussed above 
\begin{equation}
\chi_{LR}=\beta (1-q_{EA})\,, \quad \chi_{eq}=\beta \int dx (1-q(x))\,.
\end{equation}
The experimental plot of the two susceptibilities is shown in fig. (\ref{Two}). They are 
equal in the high-temperature phase while they differ in the low-temperature phase.

The experimental data are in clear qualitative agreement with the theoretical predictions, especially if we consider that some of the data are also in the critical region, where the critical exponents are far from those of mean-field theory. I think that these data provide a quite conclusive argument for the relevance of mean-field theory for the experiments.

\subsection{Spin glass simulations}

How to measure directly $q(x)$ in simulations? 

This can be done by measuring the probability distribution of the overlap:
\begin{equation}
P(q)=\overline{P_J(q)}\,,
\end{equation}
where the overline denotes the average over the different instances of the system.
To this end, we can introduce two clones of the same system that we call $\sigma$ and $\tau$.

Let us consider a system of size equal to $L^D$, that we can thermalize by numerical  simulations. We can measure the probability distribution of the clones of the same system. We define the instantaneous overlap of the two clones as
\begin{equation}
q={\sum_i \sigma_i \tau_i \over V}\,.
\end{equation}
For each sample $J$ we can measure the probability distribution $P_J(q)$. We finally have:
\begin{equation} 
x_J(q)=\int_0^q dq' P_J(q')\,, \quad \overline{x_J(q)}= x(q)\,.
\end{equation}

In other words, the probability distribution of $q$ for a given system gives the function $P_J(q)$. We must now average the function $P_J(q)$ over a large number of instances of the systems (at least one thousand if we want not too large errors) to get average probability $P(q)$. 

The computations are numerically hard because the computer time need for thermalization increases very fast with the side of the system $L$. For example in three dimensions ($D=3$) in the low-temperature phase, the thermalization time increases like $L^{z}$ with $z\approx 12$ \cite{JTIME} if we use the parallel tempering algorithm. Using standard Monte Carlo simulations the thermalization time would be exponentially large in $L$ to the appropriate power
.

\begin{figure}
\includegraphics[width=0.99\columnwidth]{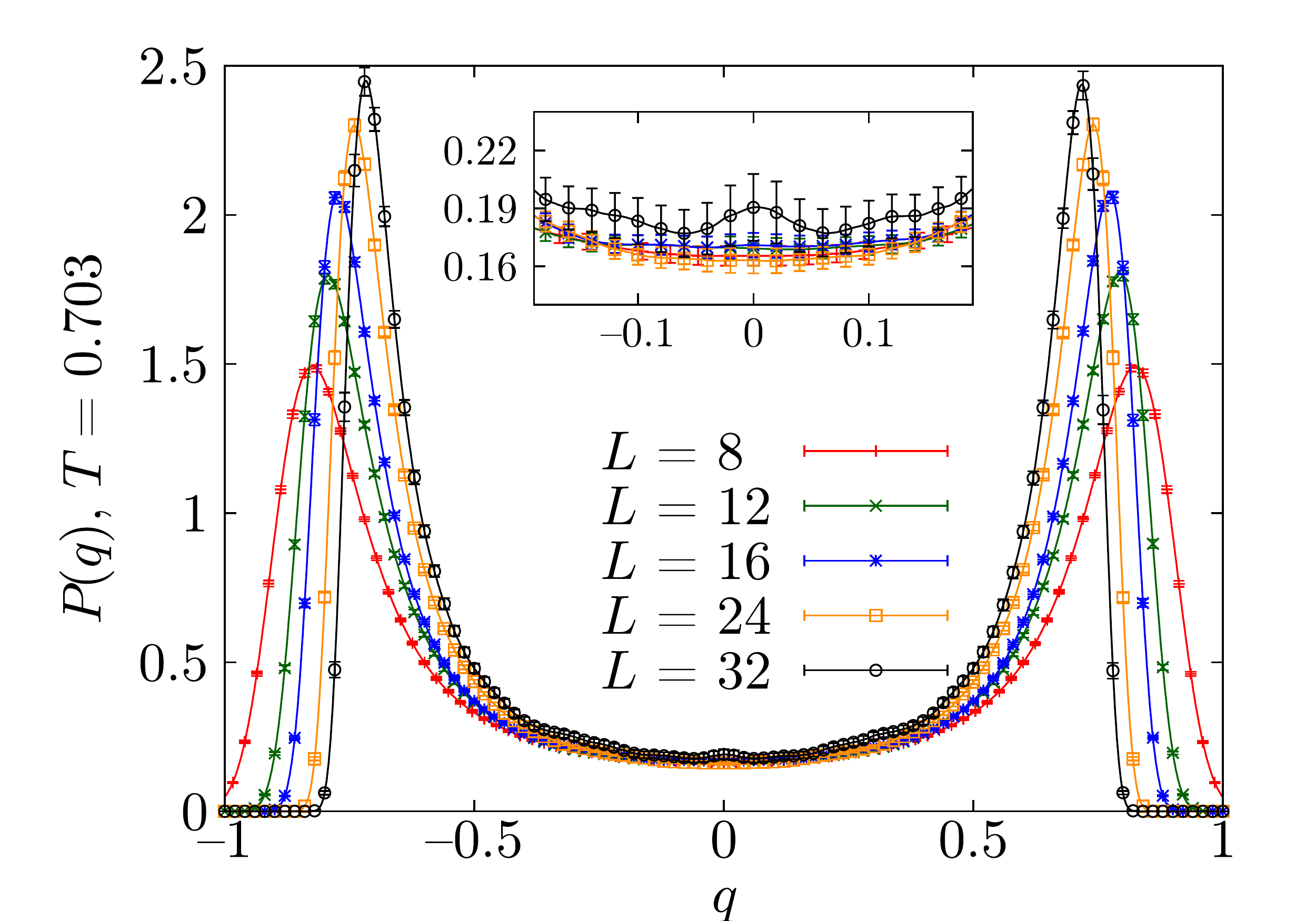}
\caption{The results from the Janus collaboration \cite{Janus} for the average function $P(Q)$ in three dimensions for $L=8,12,16,24,32$ in the low temperature regions.}\label {PQJ}
\end{figure}

\begin{figure}
\includegraphics[width=0.49\columnwidth]{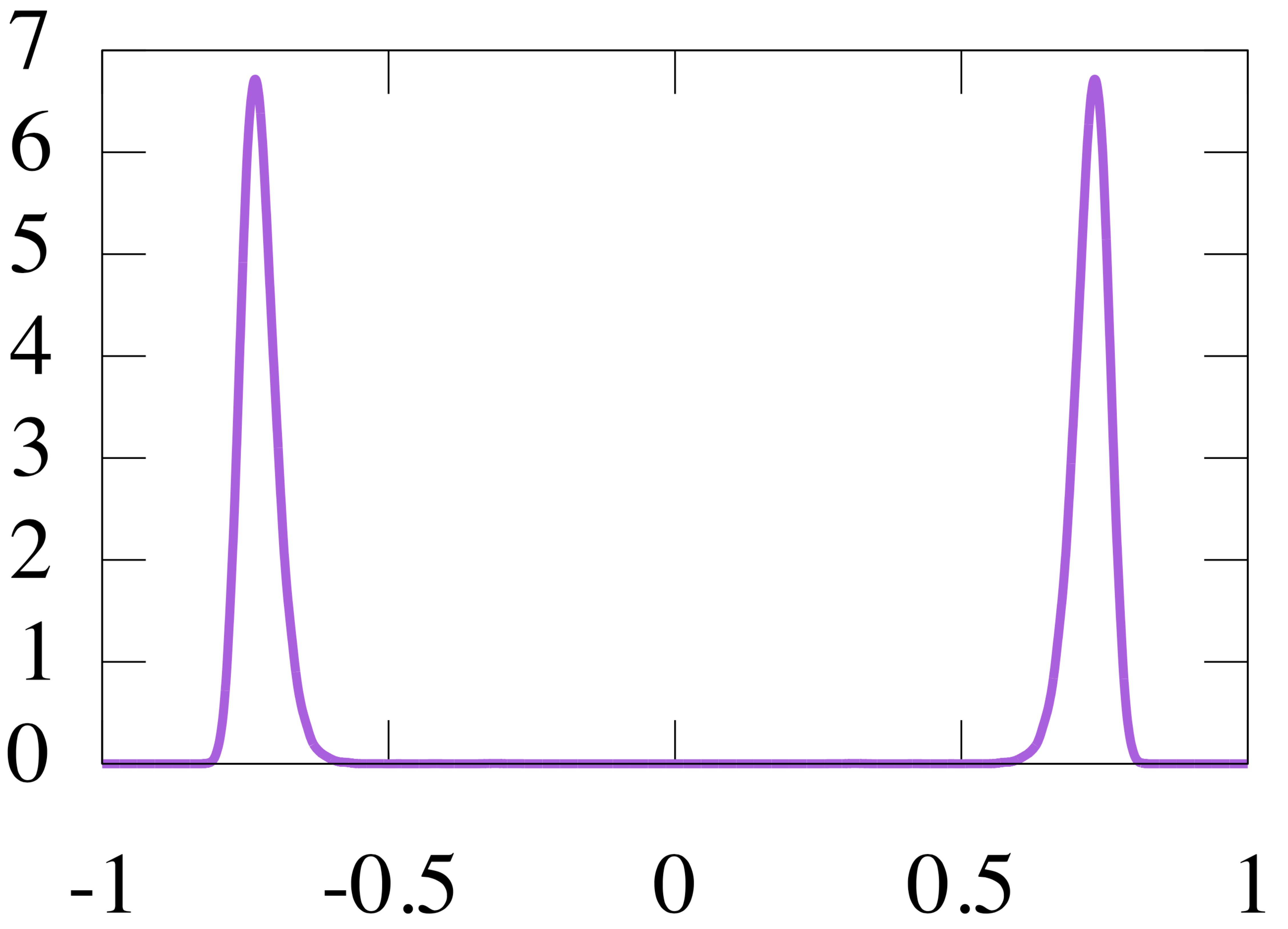}
\includegraphics[width=0.49\columnwidth]{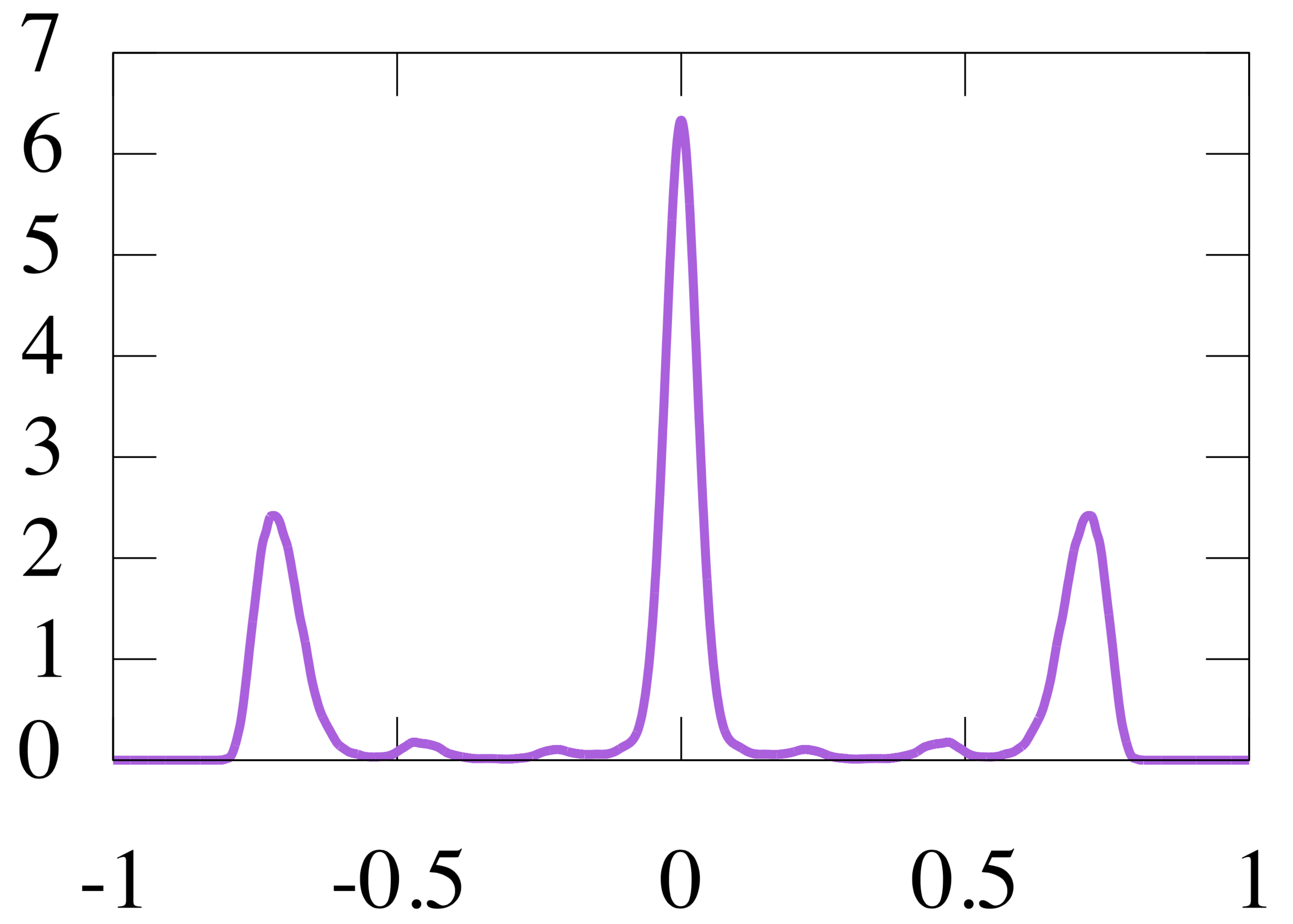}
\includegraphics[width=0.49\columnwidth]{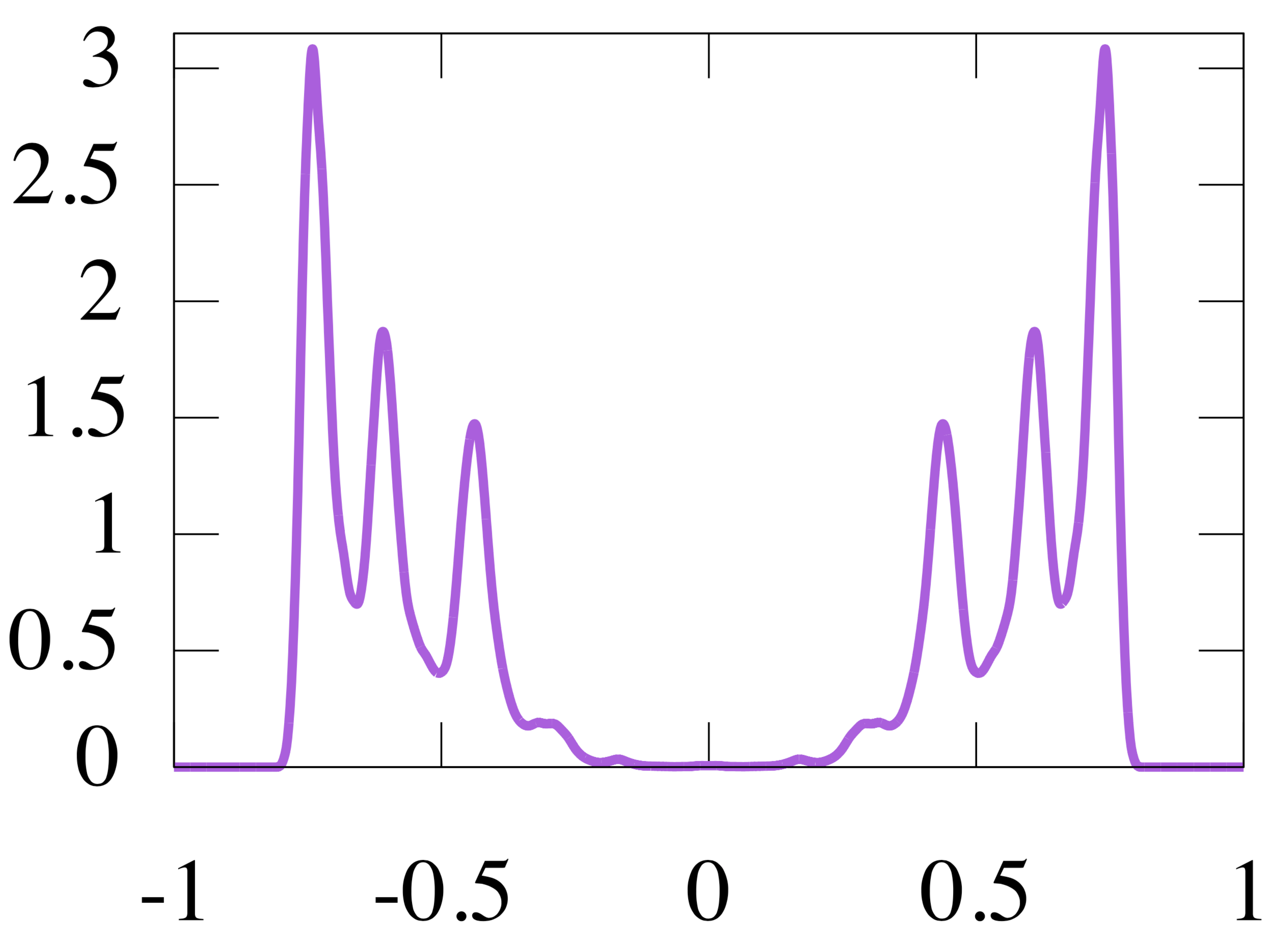}
\includegraphics[width=0.49\columnwidth]{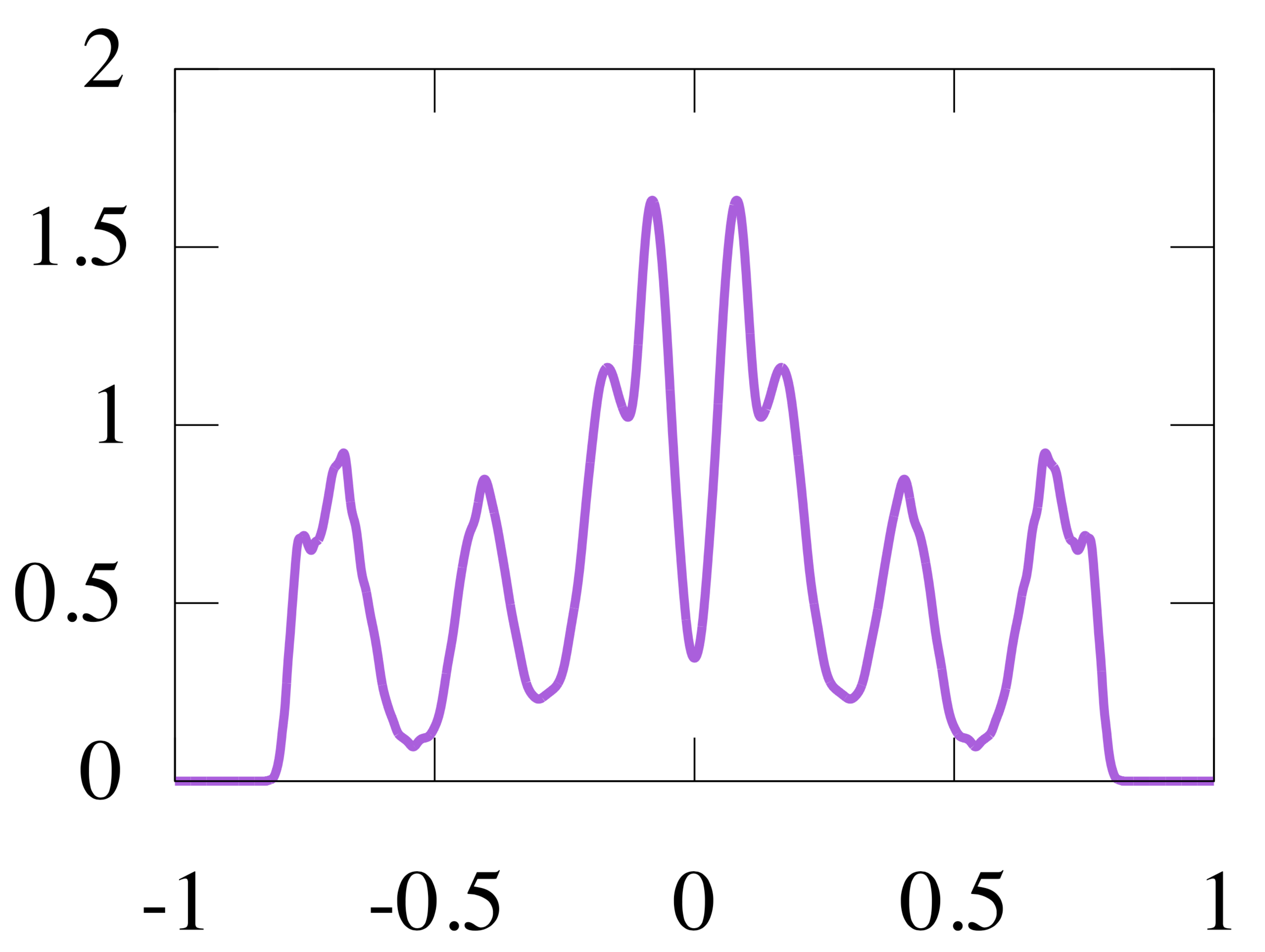}
\caption{Four different examples of the $P_J(q)$ function in three dimensions for $L=32$ corresponding to different samples in the low temperature region \cite{Janus}. The average over one thousand different samples gives the smooth function $P(q)$ of fig. (\ref{PQJ}).}\label {Example}
\end{figure}

The signature of replica symmetry breaking 
is that in the infinite volume limit the function $P(q)$ is different from two delta functions at $q=\pm q_{EA}$. 
In fig. (\ref{PQJ}) we show the results for the average function $P(q)$ from the Janus collaboration \cite{Janus} in three dimensions for $L=8,16,24,32$: the plateaux at $q=0$ has no tendency to disappear by increasing $L$. These smooth curves are the average of $P_J(q)$ functions that are very different from one another (see fig. (\ref{Example})): each sample has its own $P_J(q)$ function.
The sample-to-sample fluctuations of $P_J(q)$ follow very well the theoretical prediction as shown in \cite{SamSam}. A comparison of finite dimensional numerical simulations with infinite range model \cite{Aspel} was done in \cite{Com0,Com1}.

Unfortunately, I cannot present more results on large-scale simulations of spin glasses. I have worked on this problem for more than 20 years: I consider numerical simulations to have the same cognitive value as experiments. 
Unfortunately, a detailed presentation of the many results would bring us too far.

\subsection{Fluctuation-dissipation relations}
We can also define and measure experimentally susceptibilities that interpolate between $\chi_{LR}$ and $\chi_{eq}$: they can be directly related to the function $q(x)$.

The definition is simple: we add a small magnetic field and we impose that the perturbed system remains at an overlap greater than or equal $q$ from the original one: the corresponding susceptibility \cite{BohBoh} is given by:
\begin{equation} 
\chi(q) =\beta\int _q^1 dq'\,x(q')
\end{equation} 
This formula reproduces the values of the two previously defined susceptibilities ($\chi_{LR}$ and $\chi_{eq}$) if we set $q= q_{EA}$ and $q=0$ respectively:
\begin{eqnarray}
\chi_{LR}=\chi(q_{EA})&=&\beta(1-q_{EA})\,, \nonumber\\
\chi_{eq}=\chi(0)&=&\beta \int dx (1-q(x))\,.
\end{eqnarray}

The experimental measure of $\chi(q)$ can be done in an off-equilibrium setting and using generalized fluctuation-dissipation relations \cite{CuKu}.

I present here a very schematic version of the Cugliando-Kurchan theory \cite{CuKu,NOI}. 

Let us consider a system that has been carried to the final temperature at time 0, we wait a time $t_W$ before measurements start. If the waiting time $t_W$ is large but finite, the system is slightly off equilibrium. We can look at the magnetic response at a large time $t$ after $t_W$.

Both times are macroscopic, much larger than the characteristic microscopic time: in the experiments, they could range from a few seconds to a few hours.

We define the correlation:
\begin{equation}
C(t,t_W)= \mbox{Av}\left( \sigma_i(t_W) \sigma_i(t_W+t)\right)\,.
\end{equation}
For large times we have the modified fluctuation-dissipation relations \cite{CuKu,NOI}:
\begin{equation}
\frac{ d \chi(t_W,t)}{dt} =-\beta X(C(t_W,t)) \frac{ d C(t_W,t)}{dt}\,,
\end{equation}
where $ \chi(t_W,t)$ is the response function at the time $t_W+t$ after that an infinitesimal magnetic field has been introduced at time $t_W$.

We can eliminate the time parametrically and consider $\chi(t_W,C)$. For very large waiting time $t_W$ we should have:
\begin{equation}
\frac{ d \chi(t_W,C)}{dC} =-\beta X(C)\,.
\end{equation}
In other words, we find that when $t_W\to\infty$ the quantity $d \chi(t_W,C)/dC$ has a well-defined limit that is equal to $X(C)$.

At the end of the day, one finds that this dynamically introduced quantity $X(C)$ must be equal to the equilibrium $x(q)$, which has the meaning of a probability:
\begin{equation}
X(C)=x(q) \biggr |_{q=C} \,, \quad \chi(C)=\chi(q) \biggr |_{q=C}\,.
\end{equation}
The interpretation of these results in terms of a modified Onsager postulate has been done in \cite{FrVi}.

These results are very important as they open an experimental window on the determination of the function $q(x)$ and give a theoretical framework to study the off-equilibrium behavior.

The theory has been confirmed in a beautiful experiment \cite{OCIO}, that was done 20 years ago:  the main results for the response function versus correlation is depicted in fig. 7. It would be extremely interesting to repeat the experiment with modern ad accurate technologies, taking advantage of the progress that has been done in the theory and in numerical simulations. 

Similar results have been obtained in very careful numerical simulations, where a direct comparison with the theory is possible because the function $P(q)$ is known from equilibrium simulations (see fig. (\ref{PQJ})). The very large time span (12 orders of magnitude) helps to put under control \cite{JFDT} systematic errors related to the infinite time extrapolation.
\begin{figure}
\includegraphics[width=0.99\columnwidth]{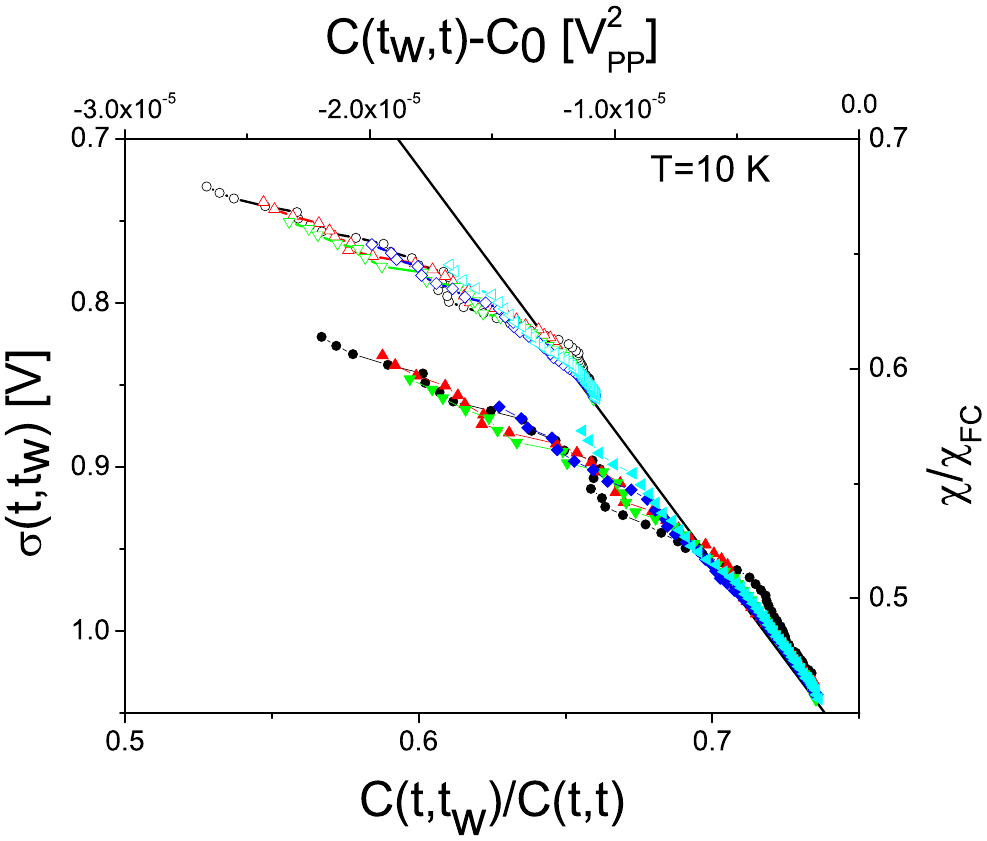}
\caption{Experimental raw results (full symbols) and extrapolations to the infinite time limit (open symbols) for  the response function (defined in eq.(30)) versus the correlation (defined in eq.(29)). The different curves span the waiting times studied: $t_w=100\ s, t_w=200\ s, t_w=500\ s,t_w=1000\ s,
t_w=2000\ s$ (from (\cite{OCIO}))
}
\end{figure}

Other impressive off-equilibrium phenomena in spin glass are memory and rejuvenation \cite{FS},
unfortunately, I cannot discuss them due to a  lack of space. Fortunately, also these effects have been partially reproduced in accurate simulations \cite{RecentJanus,RecentJanus1} where one can obtain much more accurate information using a very wide range of waiting times.

\subsection{Granular Material and hard spheres}

Classical granular matter \cite{WYART0,WYART1,WYART2,LN,9} is a problem of high interest in experimental and theoretical physics. If we neglect friction and the objects are spherical it reduces to the hard-sphere model which has been extensively studied. It was well known that by increasing the density (or the pressure) one enters a glassy phase, where nothing was supposed to happen by subsequent compression. This phenomenon corresponds to the appearance of a one-step replica symmetry breaking.

It was a real surprise when it was discovered that the analytic solution of the hard spheres thermodynamics in the mean field approximation \cite{LARGED1,FINAL,BOOKPUZ} predicted the existence at high pressure of a transition (the Gardner transition) to a marginal stable phase starting from the region where one replica symmetry was broken at one step according to \cite{KWT}. 

Approaching this transition by increasing the pressure leads to a divergent correlation time and to a divergent correlation length. 
This possibility of a transition from one-step replica symmetry breaking to a continuous replica symmetry breaking, with the consequent appearance of a marginal stable phase was discussed firstly in \cite{GKS} in the framework of a spin glass type model, but the transition was identified in 
\cite{GA}, where its properties were computed. When we cool a glass, it enters a non-equilibrium region and the possibility of a transition to continuous replica symmetry breaking in this off-equilibrium region was first discussed in \cite{P1,P2}.

This unexpected prediction was confirmed in detailed numerical analyses \cite{16,37}. This
marginal phase predicted by the replica theory of glasses,
has been directly observed experimentally in a slowly
densifying colloidal glass \cite{Lexp,Jin} and in two dimensional hard disks \cite{SD,2D}.

A spectacular result was the computation of the mean-field exponents for the jamming transition \cite{LARGED1,FINAL,BOOKPUZ} that happens in the phase where replica symmetry is spontaneously broken. For example in the mean field approximation at jamming the correction function $g(r)$ of hard spheres of diameter 1, at distance $r$ slightly greater than 1 behaves as
\begin{equation}
g(r)\propto \frac{1}{(r-1)^\gamma}\,, \quad \gamma= 0.41269 \cdots \,,
\end{equation}
where the value of $\gamma$ is obtained by solving non-linear equations.

This prediction is correct not only in high dimensions (as it should be), but it has been verified also in three and quite likely in two dimensions where some logarithmic corrections may be present \cite{TANTI}.

\subsection{Random Laser}

The theoretical interest in random lasers in connection with replica symmetry breaking started in 2006 \cite{ACRZ}. Fortunately, experimental evidence of replica symmetry breaking has been provided \cite{30,33,99}. In random lasers is possible to observe directly the occupancy of different harmonic modes and therefore one can measure directly the $P_J(q)$ function. 

Many different kinds of lasers have since been studied: not only the standard solid disordered lasers but also random fiber lasers \cite{A,B}, random laser suspensions in very viscous liquids \cite{C} heterogeneous random lasers in highly porous fibers \cite{D}.

Remarkably, similar phenomena are present also in nonlinear optical propagation through
photorefractive disordered waveguides \cite{90}. 

\section{The spin glass cornucopia}
In 1988 P.W. Anderson published seven columns in "Physics Today" discussing various issues on spin glasses. In one of the columns he described spin glasses as an amazing cornucopia \cite{PWA}: {\sl
To me, the key result here is the beautiful revelation of the structure
of the randomly "rugged landscape" that underlies many complex optimization problems (..) Physical spin glasses and the
SK model are only a jumping-off
point for an amazing cornucopia of
wide-ranging applications of the
same kind of thinking.}

Anderson was right. Here I will try to sketch some of them.

We have seen many developments in physics. I will mention here only a few examples.
\begin{itemize}
\item Structural glasses: replica symmetry breaking is relevant for the study of 
the glass transition. 

A very important step forward was done in the eighties using the mode coupling theories \cite{ModeCou}. However, it was realized that the same kind of equations can be obtained in the framework of generalized spin glass models \cite{KWT,BB}. 
This new approach was complemented by the discovery of the new replica-based thermodynamical potentials \cite{FP}. In this way, it was possible to identify the mode-coupling transition as a dynamic transition, where the correlation time goes to infinity without any thermodynamics counterpart, and to understand the behavior at low temperatures.

It was immediately clear that in reality, the dynamical transition is an artifact of mean-field theory and it is a cross-over region. In this framework, one can see the emergence of a large dynamical correlation length \cite{FP3} that has an experimental counterpart in the large non-linear susceptibility \cite{NONLIN}. 

The discovery of nearly analytically soluble non-disordered models, where replica computations could be successfully done \cite{RIT}, was very important (at least to me) for showing the relevance of the replica approach to structural glasses. Finally, first principle computations of the glass transition and the glassy properties in the low-temperature region have been successfully done \cite{MPMOL,BV}. 

Saddle points in the free energy landscape control the tunneling among different minima of the potentials, as can be seen also in mean field models \cite{SELLE5}.

Beyond mean-field theory, the dynamic transition corresponds to the crossover from dynamics dominated by narrow flat directions to dynamics dominated by barriers It was also possible to simplify the theory in such a way as to compute the critical exponents in high dimensions \cite{PRR}.

\item 
Very important progress have been made in off-equilibrium theory, the approach is quite complex because we lack Boltzmann-Gibbs statistical mechanics.
As stressed in many papers (e.g. in \cite{BOH}), the situation is different for systems that are only slightly out of equilibrium. For example, there are systems that within the experimental timescale cannot reach equilibrium because of high free-energy barriers (that may be of energetic or entropic nature): this situation typically applies to disordered systems, such as spin glasses and structural glasses. 

These systems approach equilibrium slowly, by jumping from one metastable state to another, and they remain out of equilibrium forever if continually perturbed by a slowly changing external field. 

In these systems, we can expect a separation, by many orders of magnitude, between the microscopic time scale of the system (for example, that represented by the vibrations of individual atoms) and the macroscopic time needed to cross the barrier (for example, changes in the structure of the system itself). 

The systems can then be considered to be essentially thermalized inside a metastable state, and so fluctuation-dissipation ideas can still be applied: the slowly changing overall state of the system is considered to be a small perturbation. As we have seen the standard fluctuation theorem cannot be applied to jumps from one equilibrium state to the other and we have to use the modified fluctuation-dissipation relations.

The extremely long time needed to thermalize a glassy system induced great interest in off-equilibrium phenomena like aging and many new ideas have been introduced. Maybe the most significant advance on  conceptual ground was the introduction of the modified fluctuation-dissipation relation on long-time scales for quasi-adiabatic dynamics, whose application to spin glasses I have discussed in the previous section \cite{CuKu}.
A general theory of off-equilibrium quasi-adiabatic dynamics has also been proposed \cite{QUASI}.

\end{itemize}

We have also seen many developments outside physics. I will mention only a few examples. 
\begin{itemize} 

\item Optimization theory was strongly affected. The replica approach enables scientists to compute analytically the properties of the optimal solution for large random instances. This was done for many problems: e.g. the random traveling salesman problem or the assignment problem. In this last problem, we have a set of $N$ cities and $N$ wells. The cost of connecting a city to a well is a random number in the interval [0-1]. One could show that the cost of the globally optimal solution in the large $N$ limit is $\pi^2/6$ plus computable $1/N$ corrections \cite{MPBIP,PARAT,SIC}.
\item Similar arguments \cite{MP,MP2} allow scientists to study constraint satisfaction problems where the task is to find a configuration that satisfies all the randomly chosen constraints: the most simple task to visualize is to color a graph with $M$ colors in such way that two nodes with the same color are not in contact. One question is how many colors are needed for a large random graph depending on the statistical properties of the graph \cite{COL}. 

The most spectacular results are the results of the 3SAT problem, i.e. a satisfaction problem with random clauses with three elements \cite{MZ,MZK,MPZ}. The 3SAT problem is the archetypical NP-complete problem.

\item There have been many applications of these ideas in a biological context. Let me mention only the folding of heteropolymers of biological interest (proteins \cite{OW,Sha,Manifold}, RNA \cite{RNA}) and the study of the many equilibria of ecological environments where many competing species coexist in a sometimes fragile equilibrium \cite{E0,E1,E2}.

\item These ideas helped scientists to develop new algorithms. The most famous one is the simulated annealing model of Kirkpatrick (the "K" of SK), Gelatt, and Vecchi \cite{SIMANN}. This algorithm evolved into the simulated tempering \cite{SIMTEMP} and the parallel tempering \cite{PARTEMP}, which is the state-of-the-art algorithm in many simulations problems.

Among algorithms in many different contexts (e.g. 3SAT) let me mention the survey propagating algorithm \cite{MPZ} and the backtracking survey propagation \cite{BP}. The same ideas have been applied to many others problems (e.g. compressed sensing \cite{ConSen}, matrix factorization \cite{MatFac}...).

\item We have seen that the ideas of Hebb on memory did materialize into a concrete model in the seminal model of Hopfield \cite{HOPFIELD}. In two fundamental papers Amit, Gutfreund, and Sompolisky \cite{AGS0,AGS} applied spin glass theoretical tools to derive analytically the properties of the Hopfield model. 

These papers were extremely important because they showed that the Hopfield model could be completely understood analytically: they were the trampoline for the analytic study of more complex and more realistic variations of the Hopfield model.

Bringing neural network theory closer to biological realism was a major drive in the subsequent work of the late Daniel Amit who wrote a very influential book: {\it Modeling Brain Function: The World of Attractor Neural }\cite{AMIT}. This was the starting point of many other analytic studies also intending to produce more realistic models relevant to the actual brain behavior and neurobiology \cite{PDG}.

This intellectual milieu was the origin of deep learning which now dominates  artificial intelligence.

\item
At the same time, there were many developments in the statistical mechanics of learning on different types of architecture. Remarkable results were the upper bounds of learning abilities: among them, I would like to recall the capacity of the standard perceptron \cite{G}, of the binary perceptron \cite{KM}, of recurrent neural networks \cite{DeGa} and of feed-forward neural networks \cite{FUZ}.

It is interesting to note that in the more complex cases, there is a region where replica symmetry is broken near the transition point from perfect to partial learning. This transition has many points in common with the jamming of hard spheres and in some cases, it is characterized by the same exponents \cite{FUZ}. \end{itemize}
\section{On complexity}

As we have seen multiple equilibria are at the root of my studies of complex systems.
Complex systems theory has thus been applied to various systems, the most interesting feature of complex systems is the existence of a large number of different equilibrium states.

In a nutshell what does not change over time (or changes irreversibly) is not complex, while a system that can take many different forms or behaviors certainly is. If we look around, look at ourselves, animals, ecosystems, the Earth, and the climate, we have complexity all around us. One of the most interesting outputs of my work was to find some of the physical tools needed to deal with complexity in the framework of systems with an energy function where the Boltzmann-Gibbs statistical does apply.

In many cases frustration is crucial: when frustration is present a given actor receives contradictory requests from the other actors, that cannot be simultaneously satisfied so that many compromises are possible and consequently many different equilibrium states \cite{PSTAR}. 
When we look at real complex systems, such as a living cell, a brain, a society, or a complete living being such as an animal, we always see that in these systems there is constant competition (frustration), but also cooperation between a very large number of elements that (depending on the case) can be proteins, neurons, or people. 
These systems are never in equilibrium, but they oscillate and fluctuate around some defined state of equilibrium \cite{IK}.

In this situation the system is flexible and malleable, it can adapt to changes in the environment by transitioning between various possible states without thereby losing identity: in other words, we sleep, wake up, etc., but we do not change identity. Switching among many different equilibrium states gives a living being the possibility of tuning his behavior in such a way as to adapt itself to changing needs of a fluctuating environment.

Systems, to remain complex, must have an internal balance,
%because what can happen is that the system stops being complex and goes into a different situation, into a state that can no longer be changed: 
frustration should be high enough to avoid the formation of a ferromagnetic order. However, as stressed in \cite{IK}, it is possible that this delicate mechanism of cooperation, competition, stimulation, and inhibition,is violated: in that case frustration decreases, the behavior of the system changes, the correlations between the subsystems change, the overall system begins to be no longer complex, and something abnormal happens. Examples are a tumor in the cell, a disease in the nervous system, and a dictatorship in society. At this point, the complexity of the system degrades and the functioning of the system, as a whole, is severely damaged or even eliminated. 

So I would like to conclude with a quote from my friend Imre Kondor, who asserts that the loss of complexity is dangerous, and recall the warning attributed to a great nineteenth-century historian, Jacob Burckhardt, who studied political and social processes in depth: {\sl The denial of complexity is the essence of tyranny} \cite{IK}.

\section* {Appendix\\ On the early history 
of the replica method}

Nicola d'Oresme was the greater scientist of the fourteenth century: he discovered, among many other things, what we call now Galilean invariance, i.e. the impossibility of detecting motion with local measurements. In other words, he stated that there are no absolute velocities, but only relative velocities, hence nothing contradicts Earth's motion around the Sun. 

Around 1355 he discovered that
\begin{equation}
\sqrt{x}=x^{1/2}\,.
\end{equation}

If we define the $x^n$ as the multiplication of $n$ factors $x$, $x^{1/2}$ does not make sense. 
However, if we want to define $x^n$ for non-integer $n$, we can assume that the equation

\begin{equation}
\left(x^n\right)^m= x^{(n\ m)}
\end{equation}
is valid also for non-integer $n$. In this way, we can compute $x^n$ for any rational $n$. Indeed if we put $n=1/2$ and $m=2$, we get
\begin{equation}
\left(x^{1/2}\right)^2= x^{(1/2 \ 2)}= x\,.
\end{equation}
This is the definition of the square root.

In modern times the first use of replicas was done by Robert Brout in 1959 \cite{Brout}. He aimed to treat the quenched disorder and to compute the average
(over the disorder) of the 
free energy, exactly the task we face in spin glasses. He wrote the equivalent of eq. (\ref{REP}). 

He used the replica method for simplifying the diagrammatics without discussing too much about the meaning of taking $n=0$. Indeed he worked in a perturbative setting. At each order of perturbation theory, the results are polynomials in $n$: he showed that the term proportional to $n^0$, i.e. the limit at $n=0$ of the polynomial, gives the needed result, i.e. the average over the disorder.
The results could also be obtained by explicit computations. For him, replicas were a tool for organizing a complex combinatorial problem.

In 1972 de Gennes used replicas for self-avoiding polymers \cite{DeGen}
extending the $O(n)$-symmetric $g \phi^4$ to $n=0$. In this way, he was able to compute the critical exponents for polymers using the standard renormalization group. Also de Gennes used replicas in perturbation theory: the whole computation could have been done using the standard perturbative expansions for polymers. However, in the replica approach, it was possible to use all the results of field theory. In this case, the replica group is $O(n)$; we have seen in the general case the replica group is $S(n)$, which is a subgroup of $O(n)$.

The idea of developing a $O(n)$ symmetric $g \phi^4$ field theory in terms of closed trajectories (polymers)
is also contained in the paper by Symanzik of 1969 in a completely different context \cite{Sy,PaStat}.

The same theory ($g \phi^4$ at $n=0$) can be also formulated as a Laplacian plus a
random (imaginary) potential model; for negative $g$ this theory corresponds to Anderson's theory for random potentials.
The same approach can be applied to study localization, where the replica group is the non-compact group $O(n,n)$ \cite{NonCon}. As well known from \cite{PaSou}, we can reformulate the theory introducing Fermions and this leads to the use of the $O(2/2)$ Fermionic group \cite{Efetov} instead of the undefined $O(0)$ group.

De Gennes's method was generalized in the seventies to the study of many other problems (e.g. polymers \cite{LUB}). This activity was done in a perturbative framework: all the computations could be done without replicas, but in a more cumbersome, error-prone way.

It is remarkable that in those years the idea of analytic continuation from integers appeared in many different contexts:
\begin{itemize}

\item In 1959 Tullio Regge \cite{Regge} introduced complex angular momentum. Here, Carlson's theorem was crucial to prove the uniqueness of the analytic continuation.

\item In 1972 Fortuin-Kasteleyn \cite{FK} introduced the random cluster model for the $q$ state Potts model: in the limit $q=1$ one obtains lattice percolation.

\item Always in the same year (1972) Giambiagi and Bollini \cite{GB}, and 't Hooft and Veltman \cite{HV} introduced dimensional regularization (non-integer dimensional spaces). 

The idea of spaces with non-integer dimension was independently at the basis of the very successful $\epsilon$ expansion for the critical exponents in $4-\epsilon$ expansion \cite{WF} in the framework of Wilson's renormalization group. Ken Wilson obtained the Nobel Prize a few years later. 

A rigorous definition of rotationally invariant Euclidean space is lacking; some suggestions have been done in \cite{ParTwo}.
\end{itemize}

Remarkably, the replica method for statistical mechanics was introduced by Roger Brout, who died before receiving the Nobel prize together with Englert and Higgs; others four Nobel Laureates (de Gennes, Anderson, Thouless, and Kosterlitz) contributed to the development of the method.

\section*{Acknowledgements}
I am very grateful to the hundreds of people that have worked with me on these subjects, but they are too many to list here.
However, I would like to thank Luca Leuzzi, Andrea Maiorano, Enzo Marinari, Marc Mézard, Federico Ricci-Tersenghi, Juan-Jesus Ruiz-Lorenzo, and Francesco Zamponi for reading the manuscript and for many useful suggestions.

\end{document}